\documentclass[aps,preprint,showpacs]{revtex4}
\usepackage{latexsym}
\usepackage{graphicx}

\begin{document}

\title{Two noncommutative parameters and regular cosmological phase transition in the
  semi-classical dilaton cosmology}
\author{Wontae Kim}\email{wtkim@sogang.ac.kr}
\affiliation{Department of Physics and Center for Quantum Spacetime, Sogang University, Seoul 121-742, Korea}
\author{Edwin J. Son}\email{eddy@sogang.ac.kr}
\affiliation{Department of Physics and Basic Research Science Institute, Sogang University, Seoul 121-742, Korea}

\date{\today}

\begin{abstract}
We study cosmological phase transitions from modified equations of
motion by introducing two noncommutative parameters in the Poisson
brackets, which describes the initial- and future-singularity-free
phase transition in the soluble semi-classical dilaton gravity with a non-vanishing
cosmological constant. Accelerated expansion and decelerated expansion 
appear alternatively, where the model contains the second accelerated
expansion. The final stage of the
universe approaches the flat spacetime independent of the initial
state of the curvature scalar as long as the product of the two
noncommutative parameters is less than one. 
Finally, we show that the initial-singularity-free condition is 
related to the second accelerated expansion of the universe.
\end{abstract}

\maketitle

\section{introduction}

The horizon problem and flatness problem in the standard cosmology
have been well appreciated by the
inflation model~\cite{guth} (for recent reviews, see \textit{e.g.}
Refs.~\cite{liddle,bran,guth1,linde}); however, some problems still remain unsolved.
One of them is the initial singularity problem, 
which is difficult to solve since the consistent quantum gravity has not yet been
established. Another one is the cosmological coincidence problem or
fine-tuning problem, where the only solution up to now may be an
anthropic principle~\cite{davies,susskind}.
Apart from these problems, we are facing to explain the 
late-time second accelerated expansion of the universe.
So, it has been claimed that the \textit{dark energy} defined by the negative
equation-of-state parameter $\omega \equiv p/\rho$ is responsible
for this accelerated expansion~\cite{cddett}, 
where $\rho$ and $p$ are the energy density and the pressure, respectively.
It can be easily seen from the Friedmann equation and the continuity equation
that the accelerated universe requires $\omega<-1/3$.
Note that the density of the dark energy is
assumed to be positive so that the pressure should be negative.
Especially, cosmologies with $\omega<-1$ have a defect of big rip
singularity or sudden future singularity that the scale or some
physical quantities become singular in a finite proper
time~\cite{ckw,cht,no,no1,not,cst}. 
Recently, authors in Ref.~\cite{ow,ow1,bow,ko1} 
showed that quantum effects can render $\omega<-1$ without any need of
introducing ghosts or phantoms so that it is possible to have
cosmologies 
where the equation of state parameter $\omega<-1$ without 
having big rip-like singularities.

The above mentioned cosmological problems have been studied
extensively for a long time.
However, they are usually hard to solve exactly 
and thus some simplified models may be considered in order
to get some clues and insights for realistic models.
Such models are exactly soluble two-dimensional
dilaton gravities~\cite{cghs,str,rst,mr1,cm,bpp,rt,kv,bc,bk}, especially aiming at
various cosmological problems~\cite{mr,gv,rey,bk:cos,ky:dg,ky:bd}.
So, it would be interesting to study whether a simplified
model can solve the problems and describe the late-time accelerated
expansion or not.
Recent works~\cite{ky:ncdc,ky:adsds,ks} show that
noncommutative fields make it possible to obtain
the transition from a decelerated universe to an accelerated
universe without a cosmological constant. 
However, in spite of some efforts to obtain the cosmological
phase transition, these models have some problems. One of them
is to encounter a future singularity in a finite proper time
unless an appropriate regular geometry is patched, or
it does not reproduce the first accelerated expansion in the early
universe.
Another interesting model in Ref.~\cite{ks} describes an
inflation in the early universe, the
decelerated expansion corresponding to FRW phase, and the late-time
second acceleration. However, an initial
curvature singularity still exists. So, we
would like to extend our previous work and 
obtain everywhere singularity-free cosmological solutions involving
inflation, decelerated phase, and late-time second acceleration, where the whole
profile is essentially similar to our universe chronologically.

For this purpose, we shall add two local counter-terms with the Polyakov
action of conformal anomaly in the semi-classical action, and
then impose some modified Poisson brackets with noncommutativity between
relevant fields. This process naturally yields modified sets of
semi-classical equations of motion involving two noncommutative
parameters, and then remarkably gives desired solutions.
In the next section, usual semi-classical equations of motion
obeying conventional Poisson algebra will be derived in a
self-contained manner, and it can be
shown that the expanding universe is forever without any phase change. In
Sec.~\ref{sec:mod}, new equations of motion are derived, which give nontrivial solutions depending
on noncommutative parameters. Consequently, it can be shown that
the initial- and future-singularity-free solutions along with 
cosmological phase transition are obtained when the product of
the two noncommutative parameters is less than one.
Finally, discussions will be given in Sec.~\ref{sec:dis}.

\section{permanent accelerated expansion of the universe}
\label{sec:conv}
We start with the following two-dimensional dilaton gravity
coupled to the conformal matter and its quantum correction,
\begin{equation}
   S = S_\mathrm{DG} + S_\mathrm{cl} + S_\mathrm{qt}. \label{action:total}
\end{equation}
The first term in the right-hand side is the well-known string
inspired dilaton gravity action written as
\begin{equation}
  S_\mathrm{DG} = \frac{1}{2\pi} \int d^2 x \sqrt{-g} e^{-2\phi} \left[
     R + 4 (\nabla \phi)^2 + 4 \lambda^2\right]. \label{action:dg}
\end{equation}
The classical matter action $S_\mathrm{cl}$ composed of $N$ conformal
fields $f_i$ and its one-parameter-family quantum correction
$S_\mathrm{qt}$ are given by
\begin{eqnarray}
  S_\mathrm{cl} &=& \frac{1}{2\pi} \int d^2 x \sqrt{-g} \left[-\frac12
     \sum_{i=1}^{N} (\nabla f_i)^2 \right], \label{action:cl} \\
  S_\mathrm{qt} &=& \frac{\kappa}{2\pi} \int \sqrt{-g} \left[ - \frac14
     R\frac{1}{\Box} R + (\gamma - 1) (\nabla\phi)^2 - \frac\gamma2
     \phi R \right], \label{action:qt}
\end{eqnarray}
respectively, where $\kappa = (N-24)\hbar/12$ and $\lambda^2$ is a
cosmological constant. The higher order of quantum corrections
beyond the one-loop is negligible in the large $N$ approximation where
$N\to\infty$ and $\hbar\to0$ so that $\kappa$ is assumed to be
positive finite constant.
Note that the local ambiguity terms in Eq.~(\ref{action:qt})
correspond to those of the Russo-Susskind-Thorlacius(RST) model
for $\gamma=1$~\cite{rst}, and the Bose-Parker-Peleg(BPP) model for
$\gamma=2$~\cite{bpp}. In this work, we will assume the regularization ambiguity constant to be $\gamma>2$.

In the conformal gauge, $ds^2 = -e^{2\rho} dx^+ dx^-$, defining new
fields as follows
\begin{eqnarray}
\chi &=& e^{-2\phi} + \kappa \left(\rho - \frac\gamma2 \phi \right),  \label{new:chi} \\ 
\Omega &=& e^{-2\phi} - \frac\kappa2 (\gamma - 2) \phi,  \label{new:Omega}
\end{eqnarray}
the total action~(\ref{action:total}) can be written as
\begin{equation}
S = \frac{1}{\pi} \int\/d^2 x \left[ -\frac1\kappa \partial_+ \chi
  \partial_- \chi + \frac1\kappa \partial_+ \Omega \partial_- \Omega
  + \lambda^2 e^{2(\chi-\Omega)/\kappa} + \frac12 \sum_{i=1}^N
  \partial_+ f_i \partial_- f_i \right], \label{action:new}
\end{equation}
and constraints are given by
\begin{equation}
\kappa t_\pm = - \frac1\kappa (\partial_\pm \chi)^2 + \partial_\pm^2
  \chi + \frac1\kappa (\partial_\pm \Omega)^2 + \frac12 \sum_{i=1}^N
  (\partial_\pm f_i)^2, \label{constraint:conf}
\end{equation}
where $t_\pm$ reflects the non-locality of the anomaly term in the
Polyakov action.
This integration function from the non-locality should be determined
by the boundary condition of the geometrical vacuum and
matter state.

Assuming a homogeneous space, the Lagrangian and the constraints are
reduced to
\begin{eqnarray}
  L &=& - \frac{1}{2\kappa} \left(\frac{d\chi}{dt}\right)^2 + \frac{1}{2\kappa}
      \left(\frac{d\Omega}{dt}\right)^2 + 2\lambda^2 e^{2(\chi-\Omega)/\kappa} + \frac14
      \sum_{i=1}^N \left(\frac{df_i}{dt}\right)^2, \label{lagrangian} \\
  \kappa t_{\pm} &=& - \frac{1}{4\kappa} \left(\frac{d\chi}{dt}\right)^2 + \frac14
      \frac{d^2\chi}{dt^2} + \frac{1}{4\kappa} \left(\frac{d\Omega}{dt}\right)^2 + \frac18
      \sum_{i=1}^N \left(\frac{df_i}{dt}\right)^2,  \label{constraint}
\end{eqnarray}
where the Lagrangian is defined by $S/L_0 = \frac{1}{\pi} \int dt L$
with $L_0=\int dx$, and $dx^\pm = dt \pm dx$.
Then, the Hamiltonian is 
\begin{equation}
  H = - \frac\kappa2 P_\chi^2 + \frac\kappa2 P_\Omega^2 - 2\lambda^2
  e^{2(\chi-\Omega)/\kappa} + \sum_{i=1}^N P_{f_i}^2,
  \label{hamiltonian}
\end{equation}
where canonical momenta are given by $P_\chi = - \frac1\kappa
d\chi/dt$, $P_\Omega = \frac1\kappa d\Omega/dt$, $P_{f_i} =
\frac12 df_i/dt$.

If we now define non-vanishing Poisson brackets as follows
\begin{equation}
  \{\Omega, P_\Omega\}_{\mathrm{PB}} = \{\chi, P_\chi\}_{\mathrm{PB}} =
  \{f_i, P_{f_i}\}_\mathrm{PB} = 1,  \label{PB:C}
\end{equation}
then Hamiltonian equations of motion~\cite{bk} $d{\cal O}/dt = \{
{\cal O}, H \}_{\mathrm{PB}}$ are
\begin{eqnarray}
  & & \frac{d\chi}{dt} = - \kappa  P_\chi, \quad \frac{d\Omega}{dt} = \kappa
    P_\Omega, \quad \frac{df_i}{dt} = 2 P_{f_i},  \label{eq:field:C} \\
  & & \frac{dP_\chi}{dt} = -\frac{dP_\Omega}{dt} =
    \frac{4\lambda^2}{\kappa} e^{2(\chi-\Omega)/\kappa}, \quad
    \frac{dP_{f_i}}{dt} = 0. \label{eq:momentum:C}
\end{eqnarray}
Taking $P_{f_i}=0$ for the sake of simplicity,
these equations can be compactly written as
\begin{eqnarray}
  & & \frac{d}{dt}(\chi+\Omega) = -\kappa \left( P_\chi - P_\Omega
    \right), \quad \frac{d}{dt}(\chi-\Omega) = -\kappa \left( P_\chi +
    P_\Omega \right), \\
  & & \frac{d}{dt} \left( P_\chi + P_\Omega \right) = 0, \quad
    \frac{d}{dt} \left( P_\chi - P_\Omega \right) =
    \frac{8\lambda^2}{\kappa} e^{2(\chi-\Omega)/\kappa},
    \label{eq:reduced:C}
\end{eqnarray}
which are easily solved as
\begin{eqnarray}
  & & \chi = \chi_0 + \kappa P_{\chi_0} t -
    \frac{\lambda^2}{\left(P_{\chi_0}-P_{\Omega_0}\right)^2}
    e^{2(\chi-\Omega)/\kappa},
    \label{sol:chi:C} \\
  & & \Omega = \Omega_0 + \kappa P_{\Omega_0} t -
    \frac{\lambda^2}{\left(P_{\chi_0}-P_{\Omega_0}\right)^2}
    e^{2(\chi-\Omega)/\kappa},  \label{sol:Omega:C}
\end{eqnarray}
where $P_{\chi_0}$, $P_{\Omega_0}$, $\chi_0$, and $\Omega_0$ are
arbitrary constants, and we assume $P_{\chi_0} \ne P_{\Omega_0}$.
Note that these semi-classical solutions~(\ref{sol:chi:C}) and
(\ref{sol:Omega:C}) from the Hamiltonian equations of
motion~(\ref{eq:field:C}) and (\ref{eq:momentum:C}) are essentially
equivalent to those of Euler-Lagrangian equations of motion from
the Lagrangian~(\ref{lagrangian}) since fields $\Omega$ and
$\chi$ are not the quantum-mechanical operators. 
This is
not the quantization of the semi-classical model~(\ref{action:total}).
In the next section, we will modify these
conventional Poisson brackets in order
to obtain the modified semi-classical equations of motion.

We now turn to the issue of the expanding universe by considering
the expansion rate of the scale factor,
\begin{equation}
 \dot{a}(\tau) = \frac{d\rho(t)}{dt} = \frac{\kappa P_{\Omega_0}(P_{\chi_0}-P_{\Omega_0})
 - 2\lambda^2 e^{2(\chi-\Omega)/\kappa}}{(P_{\chi_0}-P_{\Omega_0})[-2e^{-2\phi}
 - \kappa(\gamma-2)/2]} + (P_{\chi_0} - P_{\Omega_0}) \ge 0,  \label{velocity:C}
\end{equation}
where the scale factor $a(\tau)$ is the function of a comoving time
$\tau$, which is defined by $ds^2 = -d\tau^2 + a^2(\tau) dx^2$ \textit{i.e.}
$d\tau = e^{\rho(t)}dt$ and $a(\tau) = e^{\rho(t)}$, and the overdot
denotes the derivative with respect to $\tau$. It is
straightforward to show that the positive expansion follows from the condition,
\begin{equation}
P_{\chi_0} - \frac{\gamma}{\gamma-2} P_{\Omega_0} \ge 0.
          \label{cond:expanding:C}
\end{equation}
As for the constraints,
substituting the solutions~(\ref{sol:chi:C}) and (\ref{sol:Omega:C})
into the constraint equation~(\ref{constraint}) gives 
\begin{equation}
  \kappa t_\pm = -\frac\kappa4 (P_{\chi_0}^2 - P_{\Omega_0}^2) -
    \frac{\lambda^2}{2} e^{2(\chi-\Omega)/\kappa},
          \label{constraint:C}
\end{equation}
where $t_\pm$ is determined by the matter state.
The curvature scalar is calculated as
\begin{equation}
  R = \frac{2\ddot{a}}{a} =
    \frac{e^{-2\phi}}{e^{-2\phi}+\kappa(\gamma-2)/4}
    \left[ 4\lambda^2 +
    \frac{e^{-2\phi}}{[e^{-2\phi}+\kappa(\gamma-2)/4]^2}
    \left(\frac{d\Omega}{dt}\right)^2 \right],
      \label{R:C}
\end{equation}
which cannot be negative since we have assumed that $\kappa$ is
positive and $\gamma>2$. The curvature scalar in two
dimensions is directly proportional to the second derivative of the
scale factor so that the universe exhibits a
permanent accelerated expansion without any decelerated
expansion. It means that it is nontrivial to obtain the 
phase transition based on the conventional Poisson brackets.
In the following section, modification of the Poisson
brackets~(\ref{PB:C}) gives different solution showing the
desired phase change without any curvature singularities.

\section{non-singular cosmology with phase transition}
\label{sec:mod}

We are going to extend the conventional (commutative) Poisson brackets to the
modified (noncommutative) Poisson brackets characterized by the two
noncommutative parameters, $\theta_1$ and $\theta_2$, which are
reminiscent of the noncommutative algebra of the
D-brane on the constant tensor field or a charged particle moving
slowly on the constant magnetic field~\cite{sw,vfg,ko}.
We are now trying to obtain modified semi-classical solutions
from the modified semi-classical equations
of motion.
Now, the noncommutative Poisson algebra are given as~\cite{ccgm,ccgm1,bpn}
\begin{eqnarray}
  & & \{ \Omega, P_\Omega \}_{\mathrm{MPB}} 
    = \{ \chi, P_\chi \}_{\mathrm{MPB}} = 1, \nonumber \\ 
  & & \{ \chi, \Omega \}_{\mathrm{MPB}} = \theta_1, \quad \{ P_\chi,
    P_\Omega \}_{\mathrm{MPB}} = \theta_2, \label{PB:NC}
\end{eqnarray}
where $\theta_1$ and $\theta_2$ are positive independent constants.
The modified semi-classical equations of motion are given
by
\begin{eqnarray}
  & & \frac{d\chi}{dt} = \{ \chi, H \}_{\mathrm{MPB}} = -\kappa P_\chi +
      \frac4\kappa \lambda^2 \theta_1 e^{2(\chi-\Omega)/\kappa}, \label{eq:field:chi:NC} \\
  & & \frac{d\Omega}{dt} =  \{ \Omega, H \}_{\mathrm{MPB}} = \kappa P_\Omega +
      \frac4\kappa \lambda^2 \theta_1 e^{2(\chi-\Omega)/\kappa}, 
      \label{eq:field:NC}\\
  & & \frac{dP_\chi}{dt} =  \{ P_\chi, H \}_{\mathrm{MPB}} = \frac4\kappa
      \lambda^2 e^{2(\chi-\Omega)/\kappa} + \theta_2\kappa P_\Omega, \label{eq:momentum:chi:NC} \\
  & & \frac{dP_\Omega}{dt} =  \{ P_\Omega, H \}_{\mathrm{MPB}} =
      -\frac4\kappa \lambda^2 e^{2(\chi-\Omega)/\kappa} + \theta_2\kappa
      P_\chi. \label{eq:momentum:NC}
\end{eqnarray}
Note that the original semi-classical equations of
motion~(\ref{eq:field:C}) and (\ref{eq:momentum:C}) are reproduced
when $\theta_1,\theta_2\to0$. In particular, if the cosmological constant
vanishes, then $\theta_1$ decouples from the equations of motion. So,
we want to consider the non-vanishing cosmological constant to extend the
previous works.
Rewriting these equations as
\begin{eqnarray}
  \frac{d}{dt}(\chi + \Omega) &=& -\kappa (P_\chi - P_\Omega) +
      \frac8\kappa \lambda^2 \theta_1 e^{2(\chi-\Omega)/\kappa}, \\
  \frac{d}{dt}(\chi - \Omega) &=& -\kappa (P_\chi + P_\Omega), \\
  \frac{d}{dt}(P_\chi + P_\Omega) &=& \theta_2\kappa (P_\chi + P_\Omega), \\
  \frac{d}{dt}(P_\chi - P_\Omega) &=& -\theta_2\kappa (P_\chi - P_\Omega)
    + \frac8\kappa \lambda^2 e^{2(\chi-\Omega)/\kappa}, \label{eq:reduced:NC}
\end{eqnarray}
we obtain the following solutions,
\begin{eqnarray}
  \chi \!=\! C_\chi \!-\! \alpha e^{\theta_2\kappa t} \!+\! \beta e^{-\theta_2\kappa
    t} \!-\! \frac{\lambda^2}{\theta_2^2\kappa\alpha} e^{-\theta_2\kappa t +
    \frac2\kappa(\chi-\Omega)} \!-\! \frac{4\lambda^2}{\theta_2^2\kappa^2} (1 \!-\!
    \theta_1\theta_2 ) e^{\frac2\kappa(C_\chi-C_\Omega)} \mathrm{Ei} \! \left(
    \!-\frac{4\alpha}{\kappa} e^{\theta_2\kappa t} \right),& &  \label{sol:chi:NC} \\
  \Omega \!=\! C_\Omega \!+\! \alpha e^{\theta_2\kappa t} \!+\! \beta e^{-\theta_2\kappa
    t} \!-\! \frac{\lambda^2}{\theta_2^2\kappa\alpha} e^{-\theta_2\kappa t +
    \frac2\kappa(\chi-\Omega)} \!-\! \frac{4\lambda^2}{\theta_2^2\kappa^2} (1 \!-\!
    \theta_1\theta_2 ) e^{\frac2\kappa(C_\chi-C_\Omega)} \mathrm{Ei} \! \left(
    \!-\frac{4\alpha}{\kappa} e^{\theta_2\kappa t} \right),& & \label{sol:Omega:NC}
\end{eqnarray}
where $\alpha$, $\beta$, $C_\chi$, and $C_\Omega$ are integration 
constants, the exponential integral function $\mathrm{Ei}(z)$ is defined as
$\mathrm{Ei}(z) = -\int_{-z}^\infty dx \ x^{-1} e^{-x}$, and $\chi-\Omega =
-2\alpha e^{\theta_2\kappa t} + C_\chi - C_\Omega$. In addition, their
conjugate momenta are obtained as
\begin{eqnarray}
  P_\chi = \theta_2 \alpha e^{\theta_2\kappa t} + \theta_2 \beta
    e^{-\theta_2\kappa t} - \frac{\lambda^2}{\theta_2\kappa\alpha}
    e^{-\theta_2\kappa t} e^{2(\chi-\Omega)/\kappa},  \label{sol:P_chi:NC} \\
  P_\Omega = \theta_2 \alpha e^{\theta_2\kappa t} - \theta_2 \beta
    e^{-\theta_2\kappa t} + \frac{\lambda^2}{\theta_2\kappa\alpha}
    e^{-\theta_2\kappa t} e^{2(\chi-\Omega)/\kappa}.  \label{sol:P_Omega:NC}
\end{eqnarray}

From the solutions~(\ref{sol:chi:NC}) and (\ref{sol:Omega:NC}), we can
obtain the expansion condition at asymptotic regions,
\begin{eqnarray}
  \dot{a} \!\!&=&\!\! \frac{2}{-4e^{-2\phi} - \kappa(\gamma-2)}
    \left[ \theta_2\kappa (\alpha e^{\theta_2\kappa } - \beta
    e^{-\theta_2\kappa t}) + \lambda^2 \left( \frac{4\theta_1}{\kappa t} +
    \frac{1}{\theta_2\alpha} e^{-\theta_2\kappa t} \right) e^{[-4\alpha
    e^{\theta_2\kappa t} + 2(C_\chi - C_\Omega)]/\kappa } \right]
    \nonumber \\
  & &\!\! - 2\theta_2\alpha e^{\theta_2\kappa t} > 0.  \label{velocity}
\end{eqnarray}
At the asymptotic future infinity and the past infinity,
$t\to\pm\infty$, the following inequalities can be derived,
\begin{equation}
  \alpha < 0, \quad \tilde{\beta} \equiv \beta - \frac{\lambda^2}{\theta_2^2\kappa\alpha}
  e^{2(C_\chi-C_\Omega)/\kappa} > 0.  \label{cond:expanding:NC}
\end{equation}
In the intermediate region, it is not easy to write down 
the condition in a simplified form. However, it will be 
shown in later that the positive expansion rate is possible without
contraction of the universe.

\begin{figure}
\includegraphics{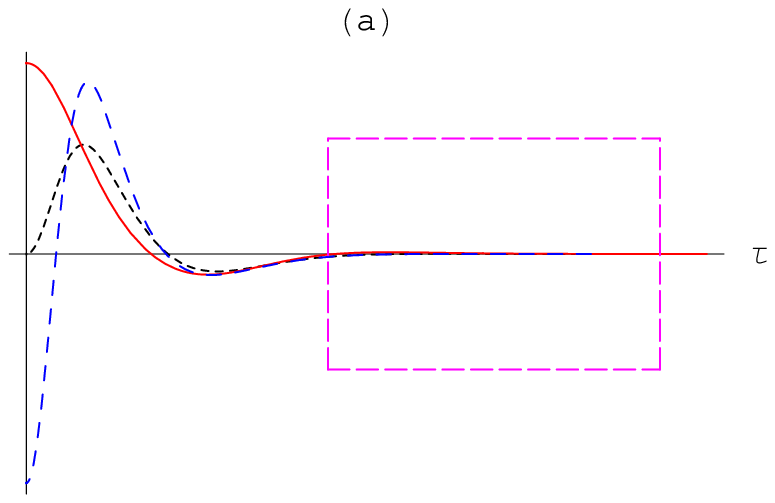}
\includegraphics{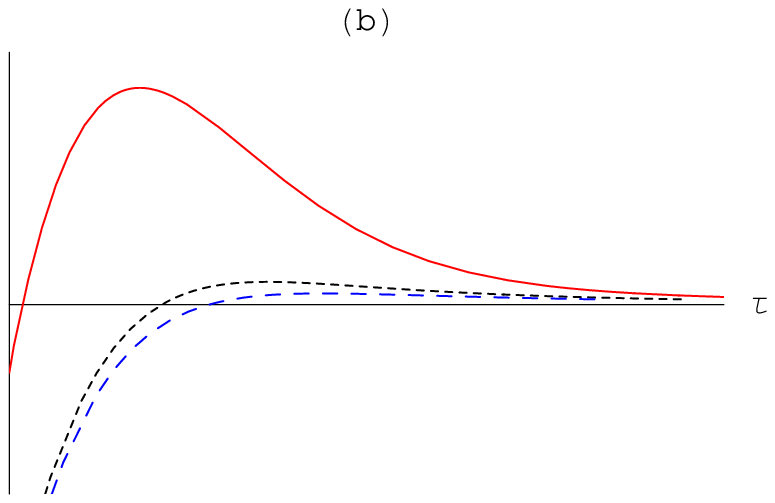}
\caption{\label{fig:curvature}
(a) The curvature scalar for $R_\mathrm{init}>0$($C_\chi=C_\Omega=0$, solid line),
$R_\mathrm{init}=0$($C_\chi=C_\Omega=4+(\gamma_\mathrm{E}+\ln5)/4$, dotted line),
and $R_\mathrm{init}<0$($C_\chi=C_\Omega=10$, dashed line) are plotted
with respect to the comoving time $\tau = \int_{-\infty}^t d\tilde{t}
e^{\rho(\tilde{t})}$.
(b) The future accelerated region in the box is magnified. The scalar
curvatures approach zero independent of their initial behaviors.
In these figures, the parameters and constants are chosen as $\kappa=1$,
$\gamma=3$, $\lambda=1$, $\theta_1=3/8$, $\theta_2=2$, $\alpha=-1$, $\beta=1$.}
\end{figure}

Now, we investigate the behavior of the curvature scalar $R$, which
is explicitly given by
\begin{eqnarray}
  R &=& -\frac{e^{-2\phi - 2(\chi-\Omega)/\kappa}}{e^{-2\phi} +
    \kappa(\gamma-2)/4} \bigg\{ \theta_2^2\kappa^2 (\alpha
    e^{\theta_2\kappa t} + \beta e^{-\theta_2\kappa t}) - 4\lambda^2
    \left( 1 + \frac{4\alpha}{\kappa} \theta_1\theta_2 e^{\theta_2\kappa t}
    + \frac{\kappa}{4\alpha} e^{-\theta_2\kappa t} \right) e^{2(\chi-\Omega)/\kappa} - \nonumber \\
  & & \qquad - \frac{e^{-2\phi}}{[e^{-2\phi} + \kappa(\gamma-2)/4]^2}
    \left[ \theta_2\kappa (\alpha e^{\theta_2\kappa t} - \beta
    e^{-\theta_2\kappa t}) + \lambda^2 \left(\frac{4\theta_1}{\kappa} +
    \frac{1}{\theta_2\alpha} e^{-\theta_2\kappa t} \right)
    e^{2(\chi-\Omega)/\kappa} \right]^2 \bigg\} - \nonumber \\
  & & - 4 \theta_2^2\kappa\alpha e^{\theta_2\kappa t} e^{-2\phi -
    2(\chi-\Omega)/\kappa}.  \label{R:NC}
\end{eqnarray}
It is interesting to note that the most leading term for $t\to-\infty$ is
$R\simeq-\theta_2\kappa C_\mathrm{cr} t$. If the following condition is met,
\begin{equation}
  C_\mathrm{cr} \equiv -4\lambda^2 (1-\theta_1\theta_2) + \frac\kappa4
  (\gamma-2)\theta_2^2\kappa^2 e^{-2(C_\chi-C_\Omega)/\kappa} = 0,
  \label{cond:sing:free}
\end{equation}
then there does not exist any initial singularity for $\theta_1\theta_2 <1$. 

\begin{figure}
\includegraphics{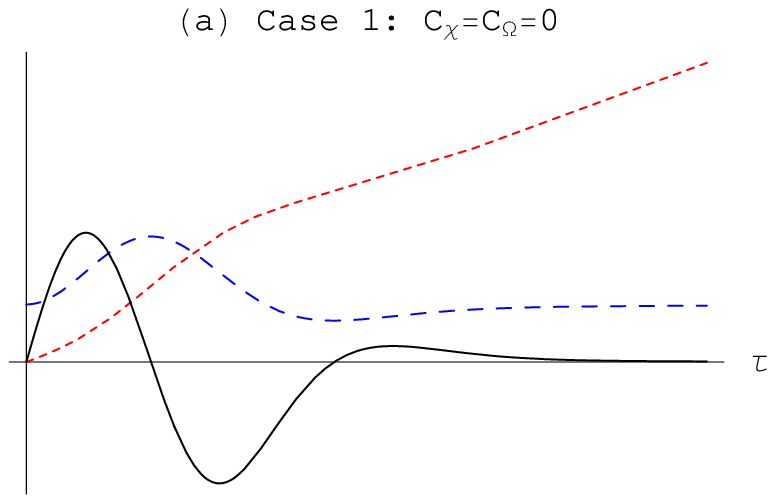}
\includegraphics{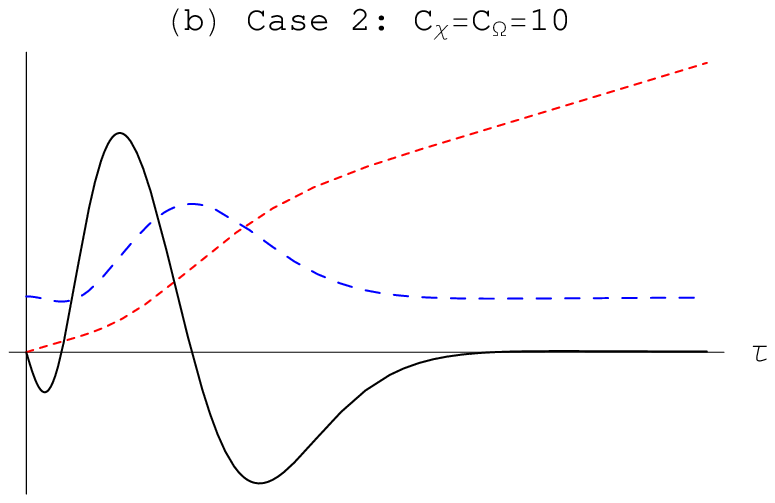}
\caption{\label{fig:scale} 
The dotted and dashed lines show the behavior of the scale factor $a(\tau)$
and the expansion rate $\dot a$ with respect to the comoving time
$\tau$, respectively.
The solid line importantly shows the profile of the acceleration $\ddot a$.
(a) Initially dS-like case($C_\chi=C_\Omega=0$) shows that the
first acceleration corresponding to the first inflation starts at the comoving
time $\tau=0$. (b)  The
first acceleration starts after a finite time for AdS like case($C_\chi=C_\Omega=10$).}
\end{figure}

We have assumed that the two noncommutative parameters $\theta_1$ and
$\theta_2$ are positive constants for
simplicity in the modified Poisson brackets~(\ref{PB:NC}).
The anomaly coefficient $\kappa$ is generically assumed to be an arbitrary
positive constant, which is related to the large $N$
limit along with the small Plank constant $\hbar$ giving the good approximation of 
the one-loop correction of matter fields~\cite{cghs,str,rst}. Especially, in this model,
the positivity is required to obtain the forward expansion of the
universe, which can be easily derived from the expansion rate~(\ref{velocity}) in the comoving coordinate, $\dot{a}\simeq\theta_2\kappa/2$ as $\tau\to0$ and $\dot{a}\simeq\kappa/(2\theta_1)$ as $\tau\to\infty$.
The ambiguity $\gamma$
is also an arbitrary constant satisfying $\gamma >2$, 
otherwise the curvature singularity appears as seen from the curvature
relations~(\ref{R:C}) and (\ref{R:NC}). Moreover, the relation of integration constants   
$C_\chi-C_\Omega$ is set to zero for simplicity, which is related to
the origin of the conformal time $t$. 
Of course, the cosmological
constant is still arbitrary in this formulation.
Assuming $\kappa=1$, $\gamma=3$, $\theta_1=3/8$, $\theta_2=2$ and 
$C_\chi=C_\Omega$, especially, $\lambda=1$, the behaviors of the curvature scalar and
the scale factor are shown in Figs.~\ref{fig:curvature} and \ref{fig:scale}, respectively.
Some different choices of constants do not change the overall profile
of these figures. 

To compare this model with the previous work $(\lambda=0)$~\cite{ks},
the crucial difference comes from the coupling of the cosmological
constant and $\theta_1$, which plays an important role as seen in Eqs.~(\ref{eq:field:chi:NC})-(\ref{eq:momentum:NC}).
In Ref.~\cite{ks}, even though the phase changing cosmological
transition has been obtained without any future singularities, the
initial curvature singularity has not been avoided.
At the first sight, some fine tuning of constants removes the
initial singularity; however, it is not the case
since the curvature singularity is a geometrically invariant quantity. 
For instance, if we take $\lambda^2=0$, then the initial
singularity free condition~(\ref{cond:sing:free}) tells us that 
$\kappa=0$ or $\gamma=2$; however, it does not remove 
singularities since the denominator in the curvature scalar~(\ref{R:NC})
may vanish~\cite{ky:adsds}.    
Therefore, in this model the nonvanishing cosmological constant along with the
noncommutativity gives the singularity free cosmological phase transition.      

From now on, we will consider the singularity-free solution
satisfying
 $C_\mathrm{cr}=0$. After some tedious calculations, the
asymptotic behaviors of the curvature scalar are written
as
\begin{equation}
  R \simeq \left\{ 
    \begin{array}{ll}
      \theta_2^2\kappa^2 C_\mathrm{N} + O(e^{\theta_2\kappa t}) &
       \textrm{ for } t \to -\infty, \\
      \frac{\kappa^2\lambda^2}{4\alpha^2} (1-\theta_1\theta_2)
       e^{-2\theta_2\kappa t} + O(e^{-3\theta_2\kappa t}) &
       \textrm{ for } t \to \infty,
    \end{array} \right. \label{curvatures}
\end{equation}
where the unbounded constant $C_\mathrm{N}$ is given by
\begin{equation}
  C_\mathrm{N} \equiv \left[ \frac\kappa4 (\gamma-2) \ln \tilde{\beta}
  - C_\Omega - \frac{4\alpha}{\kappa} \tilde{\beta} \right]
  e^{-2(C_\chi-C_\Omega)/\kappa} -
  \frac{4\lambda^2}{\theta_2^2\kappa^2} \left[ 1 -
  (1-\theta_1\theta_2) \left( \gamma_\mathrm{E} +
  \ln\left(-\frac{4\alpha}{\kappa}\right) \right) \right],
  \label{cond:zero:curv}
\end{equation}
and the Euler's constant $\gamma_\mathrm{E} = \lim_{N\to\infty} \left[
\sum_{n=1}^N \frac1n - \ln(N) \right] \simeq 0.5772$.
Since the curvature scalar is finite, eventually it is everywhere
singularity-free. In addition, the curvature scalar approaches
zero universally, which might be
similar to the attractor mechanism in the nonlinear dynamics
since the asymptotic curvature scalars are independent of $C_\mathrm{N}$
which determines the initial state of the curvature scalars.
Note that the initial state of the universe is dS-like for $C_\mathrm{N} > 0$ or
AdS-like for $C_\mathrm{N} < 0$.
The profile of the curvature scalar is plotted for the case of
$R_\mathrm{init}>0$($C_\mathrm{N}>0$),
$R_\mathrm{init}=0$($C_\mathrm{N}=0$), and
$R_\mathrm{init}<0$($C_\mathrm{N}<0$) in Fig.~\ref{fig:curvature}.
Moreover, $\theta_1 \theta_2 <1$ from the initial-singularity-free
condition is related to the late time second accelerated expansion 
since the curvature scalar of the universe should approach positive zero 
as shown in Eq.~(\ref{curvatures}). 

Next, let us remind that some of dark-energy-dominant models
have a defect called a big rip singularity that the scale blows up in a
finite time~\cite{ckw,cht,no,no1,not}. The present model is different from the
previous models, and we investigate whether this kind of singularity
appears or not. The scale $a(\tau) = e^\rho$ is expanded as
\begin{equation}
  a(\tau) \simeq \left\{ 
    \begin{array}{ll}
      \frac12 \theta_2\kappa\tau \left[ 1 + \frac{1}{12}
       \theta_2^2\kappa^2 C_\mathrm{N} \tau^2 + O(\tau^3)
       \right] & \textrm{ for } \tau \to 0, \\
      \frac12 \theta_2\kappa\tau \left[ 1 +
       \frac{1-\theta_1\theta_2}{\theta_1^2\theta_2^2\lambda^2}
       \tau^{-2} + O(\tau^{-4}) \right] & \textrm{ for } \tau \to \infty,
    \end{array} \right.  \label{scale:appx}
\end{equation}
with respect to the comoving time $\tau = \int_{-\infty}^t d\tilde{t}
e^{\rho(\tilde{t})}$. Then, we see that it is definitely finite at a
finite comoving time. It means that our model does not
have any sudden future singularities including a big rip singularity.

Let us now study the most intriguing issue of the
late-time acceleration. 
The acceleration is calculated as
\begin{equation}
  \ddot{a}(\tau) \simeq \left\{
    \begin{array}{ll}
      \frac14 \theta_2^3\kappa^3 C_\mathrm{N} \tau + O(\tau^3) &
       \textrm{ for } \tau \to 0, \\
      \frac{\kappa(1-\theta_1\theta_2)}{\theta_1^2\theta_2\lambda^2}
       \tau^{-3} + O(\tau^{-5}) & \textrm{ for } \tau \to \infty,
    \end{array} \right.  \label{acc:appx}
\end{equation}
where it vanishes at both ends.
In the intermediate 
region, the profile of the acceleration is plotted 
in Fig.~\ref{fig:scale}. It shows that the universe starts from the
inflationary era for dS-like universe($C_\mathrm{N}>0$)
while the inflation appears after a finite time $\tau$ for 
AdS-like universe($C_\mathrm{N}<0$). The former case seems to be 
more realistic. Accelerated expansion and decelerated expansion 
(FRW phase) appear alternatively,
and then it ends up with the second accelerated
expansion. The final stage of the
universe approaches the flat spacetime as long as the product of the two
noncommutative parameters is less than one. 

Next, in order to discuss the equation-of-state parameter, 
the energy-momentum tensors should be identified with the source of
the constraint equation~(\ref{constraint})~\cite{ky:ncdc,ks}, then
\begin{eqnarray}
  T_{\pm\pm}&=&-\kappa t_\pm \nonumber \\ 
    &=& \theta_2^2\kappa\alpha\beta +
  \frac{\theta_2^2\kappa^2}{4} \left(\alpha e^{\theta_2\kappa t} -
  \beta e^{-\theta_2\kappa t}\right) + \lambda^2
    \frac{\kappa}{4\alpha} e^{-\theta_2\kappa t} e^{2(\chi-\Omega)/\kappa}.
  \label{constraint:NC}
\end{eqnarray}
The energy density $\varepsilon$ and the pressure $p$ in the
comoving coordinates are defined by
\begin{eqnarray}
\varepsilon &=& T_{\tau\tau} = e^{-2\rho} \left[
  T_{++} + 2T_{+-} + T_{--}
  \right], \label{density}  \\
p &=& T_{xx} / a^2(\tau) = e^{-2\rho} \left[ T_{++} -
  2T_{+-} + T \right].  \label{pressure}
\end{eqnarray}
Because of $T_{+-} = 0$ from the equation of motion, the density and the pressure have the same
form so that the equation-of-state parameter is simply
\begin{equation}
  \omega = p/\varepsilon = 1. \label{eos:NC}
\end{equation}
This is the same with the case of the homogeneous massless conformal fields so that the
source is an ordinary matter.

Note that in spite of the plausibility of the model, it is a two dimensional toy model so
that one might wonder whether such desired behaviors persist in
four dimensions or not. The singularity free phase transition appears
when we consider the dilaton gravity and the noncommutativity
together. The action~(\ref{action:dg}) is the s-wave sector of the
higher-dimensional low-energy dilaton gravity~\cite{klp,ln} and the
quantum-mechanically induced Polyakov action~(\ref{action:qt}) is the s-wave
sector of the bosonized four-dimensional fermionic matter~\cite{as}.  
In some sense, the main body of the present model is close to the
s-wave sector of the four-dimensional model. Therefore, 
this model is expected to be partially connected with the higher-dimensional
cosmology, although the technical difficulties 
may arise from the nonlinearity
of the higher-dimensional gravity.

On the other hand, it seems that the noncommutative parameters play some important
role even when the universe is large. One should expect that they could
play a role only when the universe is small. 
In this model, unfortunately, 
the curvature scalar and the acceleration definitely depend on the
noncommutative parameters; however, the space time is flat and the
acceleration approaches zero at the
infinity, and it means that they are independent of the parameters. 
The most leading term of the late time scale in Eq.~(\ref{scale:appx}) seems to
depend on $\theta_2$ because of $a(\tau) \approx \theta_2\kappa\tau/2$;
however, it can be absorbed by redefinition $\theta_2 \tau \to\tau$.
In this process, asymptotic behaviors of the curvature scalar and the
acceleration have been unchanged at $\tau\to\infty$.
Although the noncommutativity does not affect the geometrical behaviors at the asymptotically 
infinite scale, it is still hard to resolve this problem in
this simplified model.

\section{discussion}
\label{sec:dis}
We have studied the singularity-free phase transition in the
semi-classically quantized dilaton gravity by assuming the
noncommutativity. The basic reason for this achievement is due to the
role of non-vanishing cosmological constant along with the two noncommutative
parameters, which is big difference from the previous
work. Especially, the
parameter $\theta_1$ couples to the cosmological constant through the
equations of motion. 
The cosmological constant with $\theta_1$ makes the initial curvature tensor
finite as seen from Eq.~(\ref{cond:sing:free}), 
while the other noncommutative parameter $\theta_2$ plays a
role of phase transition as seen in Ref.~\cite{ks}. 
On the other hand, we have regarded the regularization ambiguity as $\gamma>2$
since we can take the two noncommutative parameters to be small.
For $\gamma=1$(RST model), the
noncommutative parameters should be satisfied with $\theta_1 \theta_2 >1$
for the initial-singularity-free expansion 
while  $\theta_1 \theta_2 =1$ for the critical case of $\gamma=2$ (BPP
model). 
Note that it is difficult to make the two noncommutative parameters
small simultaneously unless $\gamma >2$.  

On the other hand, one might think that the noncommutative
parameters play a role to the non-singular phase transition. 
Even if this kind of phase transition seems to be interesting, however, the origin of the
noncommutativity is still unclear. In a quantum-mechanical point of view, for
instance, the
noncommutativity is related to the constraint problem so that
the Poisson brackets for a 
slowly moving unit charged particle on the constant magnetic field are given by
$\{x^i, x^j\}=-2/B\epsilon^{ij}$, $\{p^i,p^j\}=-B/2$, and
$\{x^i, p^j\}=\delta_{ij}$. 
In this work, especially for
$\gamma=2$ corresponding to the condition of $\theta_1 \theta_2 =1$,
the modified Poisson brackets are the same with the point particle
case as long as we identify $\theta_1=2/B$
and $\theta_2=B/2$, which means that the present toy model for 
$\theta_1 \theta_2  <1$ suggests that some kind complicated constraint analysis should be involved.

Final comment is in order.
After lots of dark energy models have been built 
for explanations of recent observations of the late-time accelerated
expansion, 
there have been many attempts to constrain the dark energy equation of
state by observational data sets~\cite{bm,cc,hm,mmot,hm1,uis,mg}. 
As for the constant equation-of-state parameter, 
there has been a good agreement in the literatures at
$-1.4<\omega<-0.85$, 
and its value approaches $\omega=-1$ for the dark energy based on the
cosmological constant. 
Now, one might think that this cosmological constraint excludes our
model
since our equation-of-state parameter~(\ref{eos:NC}) was fixed at
$\omega=1$. However, this is a dimensional result. Moreover, our 
two-dimensional toy model is not a realistic one while the cosmological constraint
on the equation of state is considered in four dimensions. 
In fact, the equation-of-state parameter should be
$\omega<-1/3$ to guarantee the positive acceleration in the four
dimensional general relativity. We hope our two dimensional model can
be extended to the realistic four dimensional one in the near future.


\begin{acknowledgments}
We would like to thank J.~J.~Oh and M.~Yoon for the helpful discussions.
This work was supported by the Korea Science and Engineering Foundation 
(KOSEF) grant funded by the Korea government(MOST) 
(R01-2007-000-20062-0). 
\end{acknowledgments}


\end{document}